\documentclass[prb,twocolumn,amssymb,floatfix,showpacs,showkeywords,amsmath]{revtex4}
\usepackage{amsfonts}
\usepackage{bm}
\usepackage{graphicx}

\newcommand{\bK}{{\bm K}}
\newcommand{\bKp}{{{\bm K}_+}}
\newcommand{\bKm}{{{\bm K}_-}}
\newcommand{\bb}{{\bm b}}

\newcommand{\Or}{\bm{O}_R}
\newcommand{\Ol}{\bm{O}_L}
\newcommand{\Ir}{\bm{I}_R}
\newcommand{\Il}{\bm{I}_L}
\newcommand{\tu}{\underline{t}}
\newcommand{\tpu}{\underline{t'}}
\newcommand{\Su}{\underline{S}}
\newcommand{\s}{\sigma}

\begin{document}

\title{The Berry phase of dislocations in graphene and valley conserving decoherence}

\author{A.~Mesaros}
\author{D.~Sadri}
\author{J.~Zaanen}

\affiliation{Instituut--Lorentz, Universiteit Leiden, P. O. Box 9506, 2300 RA Leiden, The Netherlands}

\begin{abstract}
We demonstrate that dislocations in the graphene lattice give rise to electron Berry phases equivalent to quantized values $\{0,\pm\frac{1}{3}\}$ in units of the flux quantum, but with an opposite sign for the two valleys. An elementary scale consideration of a graphene Aharonov--Bohm ring equipped with valley filters on both terminals, encircling a dislocation, says that in the regime where the intervalley mean free path is large compared to the intravalley phase coherence length, such that the valley quantum numbers can be regarded as conserved on the relevant scale, the coherent valley--polarized currents sensitive to the topological phases have to traverse the device many times before both valleys contribute, and this is not possible at intermediate temperatures where the latter length becomes of order of the device size, thus leading to an apparent violation of the basic law of linear transport that magnetoconductance is even in the applied flux. We discuss this discrepancy in the Feynman path picture of dephasing, when addressing the transition from quantum to classical dissipative transport. We also investigate this device in the scattering matrix formalism, accounting for the effects of decoherence by the B\"uttiker dephasing voltage probe type model which conserves the valleys, where the magnetoconductance remains even in the flux, also when different decoherence times are allowed for the individual, time reversal connected, valleys.
\end{abstract}

\pacs{72.10.Fk,73.23.-b,73.63.-b,03.65.Yz}
\keywords{}
\maketitle

\section{Introduction}

Since electrical conductance is even under time reversal, it has to be that magnetoconductance is an even function of the applied magnetic field that breaks time reversal invariance. This elementary Casimir--Onsager relation requires equilibrium conditions such that the transport is in the linear response regime.~\cite{Casimir,Onsager,CasOns,Buttiker}

Here we will present an example suggesting that in the case of finite temperature quantum transport, linear response might run into a singular limit: although the external conditions are perfectly within linear response, the parts of the current that are governed by quantum mechanics cannot equilibrate in a true sense because some quantum number is effectively conserved, with the net effect that these coherent currents feel an ``arrow of time'' negating the Onsager relations associated with true equilibrium. This might be a more general truth, but we will limit ourselves here to the specific case of graphene where we have to employ a whole array of properties specific to graphene, to come up with a design that might exhibit the aforementioned effect. We stumbled on this story in trying to find out how to turn a topological phase, that is most natural to graphene Dirac electrons, into an observable quantity.

The effect of topology on electronic properties is tied to the topological features of the underlying atomic lattice. These are the dislocations and disclinations. Although disclinations are rather unnatural according to the standard theory of elasticity~(or plasticity),~\cite{Kleinert} the global influences they exert on graphene's Dirac electrons~\cite{LCfictflux,fullerene} have been relatively thoroughly studied, with a special focus on the similarities with the holonomy structure of fundamental Dirac electrons in a curved space--time.~\cite{Novoselov,Vozmediano,Osipov,Furtado,Sitenko}

However, dislocations have been largely ignored,~\cite{ivscatt,eeint} although these are ubiquitous topological defects in any solid. In contrast to disclinations they require only finite energies to be created, so that it is virtually impossible to prepare a crystal that contains no dislocations at all. These have not been found in the graphene flakes produced by the Manchester method,~\cite{Geim} likely because dislocated graphene does not survive this rather violent method of preparing a sample. With more sophisticated manufacturing methods it is expected that graphene dislocations will be abundant.~\cite{atomicdefects}

A dislocation, due to its topological nature, exerts influence also far away from the core. The question arises as to what happens to a quantum--coherent graphene Dirac electron that is transported around it. We analyze this problem in Section~\ref{secII}; the outcome is a holonomy structure of pleasing simplicity. The dislocation is the topological defect associated with translations,~\cite{Kleinert} and since translations are Abelian, the holonomy is akin to the holonomy associated with electromagnetism --- the Aharonov---Bohm~(AB) phase. A crucial difference is that the Dirac electrons feel a (pseudo--)flux of the same magnitude, but opposite sign in the two valleys, which is a consequence of dislocations leaving the system time reversal invariant. In addition, the topological charge of the dislocation appears in quantized units corresponding to fractions $\{0,\pm\frac{1}{3}\}$ of the magnetic flux quantum.

\begin{figure}[t]
\includegraphics[width=0.8\linewidth]{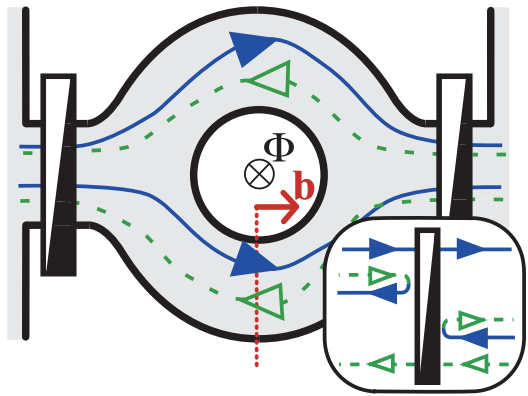}
\caption{The modified graphene Aharonov--Bohm device. This is the usual ring pierced by an external magnetic flux $\Phi$, but now with a dislocation with Burgers vector ${\bm b}$ in the center acting as a pseudoflux on the electrons with a definite valley number. Both leads are equipped with a valley polarizer: ideally these transmit fully, say, electrons in the ${{\bm K}_+}$ mode~(solid lines) moving from left to right, while ${{\bm K}_-}$ electrons~(dashed lines) propagating in the same direction are reflected to ${{\bm K}_+}$ mode moving in the opposite direction, as required by time reversal invariance. In Sec.~\ref{secIIIA} it is argued that at temperatures such that the device size is of order of the phase coherence length, only Feynman paths traversing the device once contribute to the magneto--oscillations: ${{\bm K}_+}$ modes moving from left to right~(solid lines) sense the direction of the Burgers vector in a way that is opposite to the ${{\bm K}_-}$ modes moving from right to left~(dashed lines), and this implies that the dislocation pseudoflux offsets the magneto--oscillations. At low temperatures the long Feynman paths explore the ``backside'' of the polarizers and B\"uttiker's law is restored.}
\label{fig1}
\end{figure}

For detection purposes, one envisages a typical AB experiment where the dislocation is put in the middle of a ring~(Fig.~\ref{fig1}). The AB oscillations are influenced by the presence of the dislocation holonomy, and we will discuss this in Section~\ref{secII'}. It turns out that the dislocation topological phase could in principle be measured, after it is disentangled from the elastic scattering of impurities by disorder averaging. Its effect is connected to the AB oscillation amplitudes, which are in practice less reliable due to the standard mesoscopic clutter of the oscillations.

If the current was carried exclusively by electrons in one valley~\cite{filter} the situation would be quite different, since these sense the dislocation Berry phase as indistinguishable from a real magnetic flux. Abstractly, it seems the dislocation Berry phase could thus cause the offset of the magneto--oscillations, which would violate the Onsager relation. We therefore consider the concrete possibility of valley filters installed at the input and output terminals of our dislocated AB ring~(Fig.~\ref{fig1}). The time reversal invariance puts a constraint on the general workings of the valley filter: when it is perfectly transparent for electrons in valley $\bKp$ coming from the left (thus completely reflecting the $\bKm$ valley), a $\bKp$ electron impinging on it from the right will be unitarily backscattered to the $\bKm$ valley. Deep in the quantum regime where the phase coherence length is large compared to the size of the ring $L$, long Feynman paths encircle the ring many times, having ample opportunity to explore the ``backside'' of the valley filters, with the effect that the quantum current equilibrates over the two valleys, restoring the evenness of the magnetoconductance. When temperature rises, the phase coherence length shrinks and becomes of order of $L$. The coherent part of the current that is sensitive to the topological phases can still be detected but now it is dominated by Feynman paths that traverse the ring only once. These can no longer explore the backside of the valley polarizers, and so can no longer sense that time reversal is unbroken, with the consequence that the magnetoconductance becomes uneven. We will address this more quantitatively in Section~\ref{secIIIA}. The simple essence of the argument is the observation that even in a linear response measurement, the quantum coherent part of the current cannot reach a true equilibrium. The underlying assumption is that the electrical currents are conserved separately for the two valleys everywhere in the device, except at the valley polarizers. Since these are separated in space by a length $L$, this current can be regarded as effectively conserved when the phase coherence length becomes of the order of $L$ for the purposes of quantum coherent phenomena that depend on the conservation of valley current. This quantum conserved current acts in analogy with the role of conservation laws in conventional hydrodynamics to prohibit the system from reaching equilibrium.

The argument as presented implicitly rests on the language of Feynman paths and there are precedents known where qualitative arguments of that kind can be quite misleading with regard to quantum transport.~\cite{QDotInRing} A superb theory describing transport deep in the quantum regime is the Landauer--B\"uttiker scattering matrix formalism and we will address the workings of our device in this language in Section~\ref{secIIIB}. It seems that the formalism is inherently static, revolving around elastic scattering which is sufficient at zero temperature, but at finite temperature the role of imaginary time becomes central in properly accounting for the effects of inelastic scattering. Among the various attempts,~\cite{DattaProbe,Paladephase} the voltage probe approach to incorporating dephasing~\cite{ButtikerPhantom,ImagandPhant} suggested by B\"uttiker is particularly prominent. It amounts to attaching an extra terminal to the coherent quantum device, with the effect of scrambling the phase of the waves entering this phantom reservoir. This has a respectable track record with regard to correctly modeling the effects of decoherence on quantum transport~(e.g., Ref.~\onlinecite{QDotInRing,ImagandPhant,exptestbutt,exptestbutt2,PhantvsImagSymm}). We straightforwardly extend this method to the present device by requiring that the dephasing reservoirs do not affect the valley quantum number, assuming the intervalley inelastic time to be infinitely long. As long as time reversal and unitarity of scattering are present, it follows generally from this formalism that magnetoconductance is even,~\cite{Buttiker} a fact in this context referred to as B\"uttiker's theorem. We prove that this holds even when different dephasing times are allowed for the two valleys, which are connected by time reversal. Furthermore, we explicate how the dislocation phase signature in the AB oscillations remains the same as in the zero temperature calculation.

We hope that this story will motivate experimentalists to realize our device in the laboratory. It appears to us that the matters at stake cannot be decided by theoretical means alone, as we will substantiate in the rest of this paper.

\section{Electron Berry phase and the Burgers vector of dislocations}
\label{secII}

\begin{figure}[h]
\includegraphics[width=0.95\linewidth]{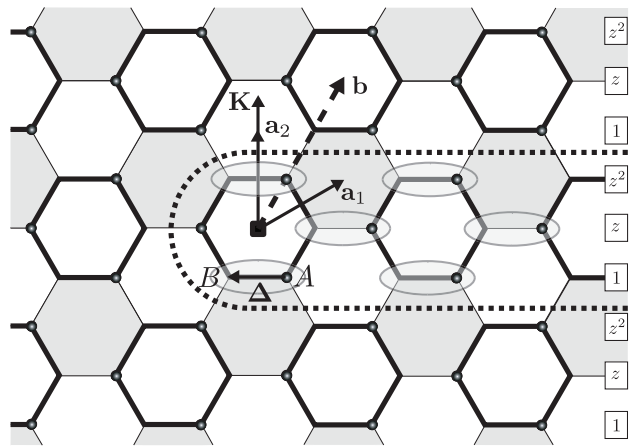}
\caption{The electronic structure and dislocations in graphene. By removing rows of unit cells a dislocation with Burgers vector $\bb$ is created. The ellipses indicate which unit cells can be removed to obtain a ``trivial'' dislocation not carrying a net topological charge, as can be seen for instance by counting the phases of the Bloch waves. An arbitrary Burgers vector starts from the central square and reaches the center of some hexagon, and this labels the dislocation's class: bold hexagon sides represent trivial dislocations ($d=0$), grey shade the $d=\frac{1}{3}$ class, and white fill the $d=-\frac{1}{3}$ class. Graphene's Dirac electrons carry unit cell~($A/B$) and ``valley'' $\bf{K}_{\pm}$ indices. The phases of the Bloch waves of the $\bKp$ states on the rows of the defect--free lattice are indicated at the right in terms of $z=exp[i 2\pi/3]$. By creating the Volterra cut associated with the dislocations it follows that the Dirac electrons experience topological phase jumps of $\frac{2\pi}{3}$ and $-\frac{2\pi}{3}$ for dislocation class $1/3$ and $-1/3$, respectively. The $\bKm$ states experience the opposite phase jump. Note that the phase jumps are independent of the $A/B$ quantum numbers, because the dislocation does not affect the intra unit--cell structure.}
\label{fig2}
\end{figure}

One of the two possible topological crystal defects, the dislocations, are omnipresent in crystals in general. A dislocation is in principle obtained by the Volterra construction as follows: a semi--infinite strip of unit cells is removed from a crystal, and the open edges are glued back together along the Volterra cut, leaving some imperfections at the original beginning of the strip~(the core), see Fig.~\ref{fig2}. Tracing a closed loop around the defect core, but drawing it in the perfect lattice, one finds a non--closure, equal to some lattice vector --- the Burgers vector. This persists for loops of arbitrary size, and so the effect of the defect on electron wavefunctions is global and long--ranged. This property enables one to model the defect as a nontrivial boundary condition on the wavefunction at the Volterra cut, which can be imposed by a gauge field in a reversal of the usual argumentation for appearance of the Aharonov--Bohm effect.~\cite{Hamilton} The difference with the case of disclinations,~\cite{fullerene,LCtopphase,Vozmediano} the other topological crystal defect, in which one cuts out a pie segment of the lattice, is that instead of rotating the electron spinor, under the influence of the translational dislocation,~\cite{Kleinert} the spinor is \textit{translated} by the Burgers vector to maintain single--valuedness.

It is known that disclinations cause a deficit angle in loops circling the core, which in graphene can be any of the five multiples of $\pm\frac{\pi}{6}$, producing a variety of physical effects,~\cite{LCtopphase} and are interesting primarily because of their occurrence in nanocones and fullerenes.~\cite{LCfictflux,fullerene}

The theoretical study of dislocations, however, has so far been scarce.  Random distributions of dislocations have been discussed from the perspective of their statistical influence on coherence and electron propagation.~\cite{ivscatt,eeint} We will here address a different set of phenomena associated with their topology. We will show that although the topological charge of a dislocation could be \textit{any} lattice vector, they act as a simple Aharonov--Bohm flux located at the defect core, of opposite signs in two valleys. They fall into three possible classes --- a trivial one~(zero flux), and two of opposite sign~($\pm\frac{1}{3}$ flux).

Let us start by reviewing the standard low--energy, continuum description of the graphene electron states coming from the $p_z$ carbon orbitals.~\cite{Slonczewski,LuttKohn,effmassgraph,Wallace} The two ``valley'' Dirac points are labeled by ${\bm K}_{\pm}=\pm{\bm K}$~(Fig.~\ref{fig2}), and the unit cell contains two atoms~(labeled $A$ and $B$), yielding a total of four massless states. In this basis the wavefunctions are described by a slowly--varying four component spinor. Operators acting on the $A$ and $B$ states without mixing the ${\bm K}_{\pm}$ valleys are written as Pauli matrices $\sigma_a$, $a\in \{ 1,2,3\}$; while the valley degeneracy is tracked by a second set of $\tau$ Pauli matrices. To lowest order this yields the usual Dirac Hamiltonian,
\begin{equation}
\label{generalDirac}
H=-i \left [ (\underline{{\bm K}}\cdot{\bm{\partial}})\tau_3\otimes\sigma_1 + (\underline{{\bm \Delta}}\cdot{\bm{\partial}})\openone\otimes\sigma_2\right ],
\end{equation}
where the energy is measured in units of $\hbar v_F$, $\underline{{\bm K}}$ is the normalized ${\bm K}$ vector and $\underline{{\bm \Delta}}$ the normalized vector connecting the $A$ and $B$ sites~(see Fig.~\ref{fig2}).

We now consider the influence of dislocations on such Dirac fermions, associated with the translation by a Burgers vector~$\bb$ at the modified boundary arising from the Volterra cut. The components of $\Psi$ are coefficients multiplying the Fermi states, ${\bm K}_\pm A/B$, being Bloch eigenstates of the crystal lattice, and a translation by a lattice vector is therefore equivalent to a multiplication by the corresponding phase factor~$exp(i{\bm K}_\pm\cdot{\bm b})$. This yields the $U(1)$ holonomy
\begin{equation}
\label{holonomy}
U({\bm b})=e^{i({\bm K}\cdot{\bm b})\tau_3}=e^{i\frac{2\pi}{3}(b_1-b_2)\tau_3},
\end{equation}
where $b_1$ and $b_2$ are the integer components of the Burgers vector ${\bm b}$ in the lattice basis~(see Fig.~\ref{fig2}). The dislocations thus separate into three equivalence classes, labeled by $d\in\{0,\frac{1}{3},-\frac{1}{3}\}$, with $3 d\equiv (b_1-b_2)\ mod\ 3$, where the period of 3 follows from the periodicity of the Fermi states~(see Fig.~\ref{fig2}). Different from the case of disclinations,~\cite{LCfictflux,fullerene} this is independent from the $A/B$  sublatitce pseudospin quantum number since translations carry no information on the structure inside the unit cell. Instead, this phase does depend on the valley quantum number in a simple way: the absolute magnitude is the same and the phases in the two valleys just differ by a minus sign.

Avoiding the dislocation core (which shrinks to a point in the continuum limit), its influence can be encoded by adding a $U(1)$ gauge coupling to the Dirac Hamiltonian in Eq.~\eqref{generalDirac},
\begin{equation}
H_{disl}=H-i\frac{\bm{\sigma}\cdot\bm{e}_\varphi}{2\pi r}({\bm K}\cdot{\bm b})\tau_3,
\end{equation}
where $r$ and $\varphi$ are the standard polar coordinates, taking the dislocation core as origin. The induced gauge field is in in precise correspondence with the one of an Aharonov--Bohm solenoid with flux $\mp d$ in units of $e/\hbar$, for the $\pm\bK$ valley electrons. Numerical simulations have already hinted that dislocations behave as pseudo--magnetic fluxes, in that they create vortex currents around their core.~\cite{zhang}

We close this subsection by discussing the role of time reversal invariance, being an important issue in this paper. Real magnetic fields break time reversal, expressed through an antiunitary operator~($\mathcal{T}$) which involves complex conjugation~(operator $C$). Time reversal applied to the graphene Dirac electrons exchanges the Fermi points and the corresponding modes in the leads. The time reversal operator can be chosen as simply $\mathcal{T}\equiv\tau_1 C$. One has $[\mathcal{T},U(\bb)]=0$, as well as $[\mathcal{T},H]=0$: time reversal amounts to flipping the external magnetic field, reversing the direction of motion of electrons, and switching them to the opposite valley, while keeping $\bb$ unchanged. After all, the lattice defect is just a complicated rearrangement in the lattice potential, and cannot break time reversal symmetry. The time reversal symmetry also dictates that the dislocation pseudoflux has to be of opposite sign for the two valleys.

\section{General properties of a dislocated AB ring}
\label{secRing}

In the previous section we have shown that dislocations correspond to quantized magnetic fluxes, carrying however opposite signs with respect to the two valleys. The standard way to measure such fluxes is by measuring the conductance of an Aharonov--Bohm ring device, as indicated in Fig.~\ref{fig6}. Besides the usual magnetic flux that can be pierced through the ring, we consider one or more dislocations located inside the ring. The electrons do not explore the dislocation cores and only communicate with the ``lines of missing atoms'' attached to the dislocation cores that cross the ring ``somewhere''~(the choice of this missing line is actually a gauge freedom on its own~\cite{Kleinert}).

With regard to the feasibility of realizing this device in the laboratory, we already argued in the introduction that dislocations should be plentiful in graphene that is produced with non violent methods. Graphene structures with a size $\lesssim 1\mu m$ have been manufactured, and show quantum transport phenomena, including AB oscillations.~\cite{Alberto} Concerning the final important ingredient, the valley polarizers, it has been suggested that the valley polarized currents could be generated using valley filters constructed from thin strips of graphene with zigzag edges.~\cite{filter}

At this point one may ask how realistic is it to assume that valley currents are conserved at the mentioned length scales. The first issue is that the intravalley inelastic scattering time should be, at a given temperature, much smaller than the intervalley inelastic scattering time, to satisfy the requirement that the intravalley phase coherence length becomes quite small while the valley polarization is not destroyed at this temperature. The origin of these inelastic times is of course not mysterious: it is rooted in Fermi--liquid electron--electron and electron--phonon scattering. Although we are not aware of unambiguous experimental information,~\cite{time1,time2,time3,time4,time5,time6} it is widely believed that the intervalley inelastic time is indeed much longer, because of the kinematical bottleneck that is active both for electron--electron and electron--phonon scattering in the form of the large momentum that has to be absorbed when the on--shell electrons are scattered between valleys. In fact, the elastic intervalley scattering is more worrisome since valley quantum number is quite fragile, being rooted eventually in lattice potentials, and one expects it to be very sensitive to the imperfections of real life devices.

There are indications from theoretical studies that the problems are manageable as long as one does not make the structures too narrow. The boundaries do not seem to play a critical role,~\cite{Anton,Adam} and there is numerical evidence for valley conservation in the ring geometry.~\cite{CarloAB} Eventually, one can contemplate even smooth terminations using mass confinement due to potentials,~\cite{Patrik} which automatically preserve the valley.

\section{Dislocated Aharonov--Bohm ring at zero temperature}
\label{secII'}

The focus of this section will be on the fully coherent quantum transport at zero temperature and in this regime the valley filters do not have a decisive influence on the conductance. As announced in the introduction this might be different at finite temperatures. The conclusion of this section will be that when the valley currents are conserved the dislocation Berry phase is observable in principle, but harder in practice: after inserting a dislocation in the ring, keeping it the same otherwise, especially with regard to point disorder, its presence can be deduced in principle from changes in the amplitude of magnetoconductance oscillations. When the intervalley scattering length becomes smaller than $L$ (ring arm length), the electron transport carries no information any longer pertaining to the presence or absence of the dislocation(s).

Let us focus on an ideal device which has ballistic transport in the arms, the magnetic field, and the dislocation. The total topological phase contribution to the wavefunction on traversing the ring is just the sum of the electromagnetic~($\Phi$, in units of $h/e$) and defect, Eq.\eqref{holonomy}, pseudofluxes, since both electromagnetism and dislocations are governed by Abelian symmetries~($U(1)$ and translations, respectively). Starting with the case when intervalley scattering is assumed to be absent, while the valley filters of Fig.~\ref{fig1} are switched off, the current is due equally to carriers from both valleys. We learn from Eq.~\eqref{holonomy} that for the nontrivial dislocations the magnetoconductance curve $G(\Phi)$ is shifted by $\frac{2\pi}{3}$ for carriers at one Dirac point, and by $-\frac{2\pi}{3}$ at the other, while the signs reverse on switching the dislocation class. Adding the two currents, each with the associated phase shift, results in the magnetoconductance $G(\Phi+\frac{2\pi}{3})+G(\Phi-\frac{2\pi}{3})$. Fourier expanding this as $G(\Phi)=G^{(0)}+G^{(1)}\cos(\Phi)+\ldots$ shows that the harmonics of order $3 n$, for $n\in\mathbb{Z}$, do not change, and all others are multiplied by a factor $-\frac{1}{2}$. In particular, the fundamental frequency oscillation~(with period $\frac{h}{e}$) is halved in amplitude. This means that the influence of the dislocation Berry phase is quantitative, affecting only the amplitudes of the Fourier components of the AB oscillations. But these are also affected by point disorder, which gives rise to the standard sample--to--sample mesoscopic fluctuation.

Let us address these matters quantitatively using the Landauer--Buttiker scattering matrix formalism.~\cite{Buttiker} We employ a model where the polarizers and two arms of the ring are described by a single scatterer each, completely analogously to the normal metal ring case in Ref.~[\onlinecite{normmetring,Buttiker1DRing}]. The modes are labeled by transversal momentum and valley, while the electrons can propagate in both directions, inside both the left and right lead.~\cite{mesotrans} The amplitudes of outgoing modes, $\bm{O}\equiv\left(\begin{smallmatrix}\Ol\\\Or\end{smallmatrix}\right)$, and incoming modes, $\bm{I}\equiv\left(\begin{smallmatrix}\Il\\\Ir\end{smallmatrix}\right)$, are connected by a scattering matrix $\Su$, with $\bm{O}=\Su\bm{I}$. The important submatrices of $\Su$ are $\tu$ and $\tpu$, given by $\Or=\tu\Il$ and $\Ol=\tpu\Ir$, where $\Il/\Ir$ are columns of amplitudes of incoming~(into a scatterer) modes from the left/right, and $\Ol/\Or$ are columns of amplitudes of outgoing modes from the left/right, see Fig.~\ref{fig6}(b); thus $\tu$ and $\tpu$ are $M\times M$ matrices, where $M$ is the number of modes in one lead. We employ the usual simplification of using only a single transversal mode~($M=2$) for simplicity, with $\Il=\left(\begin{smallmatrix}\Il^{+}\\\Il^{-}\end{smallmatrix}\right)$, etc., with the expectation that the salient features of this model survive in the realistic case of graphene with more modes. In the remainder the $\pm\bK$ modes are labelled by $\sigma\in\{+,-\}$, and we follow the convention that the scattering matrices are defined by organizing the amplitudes in columns as described above. Notice that since $\bKm=-\bKp$, time reversal exchanges the two valleys, connecting e.g. incoming~(left moving) electron amplitude in one valley, to the outgoing~(right moving) amplitude in the opposite valley, on the same side of the scatterer.

The scattering matrices used to calculate the total $\Su$, are as follows: the splitter~(circle in Fig.~\ref{fig6}(a)) has a perfect transmission and divides the amplitude equally between the two ring arms, corresponding to the leads strongly coupled to the ring, i.e. $\epsilon=\frac{1}{2}$ in Ref.~\onlinecite{Buttiker1DRing}; the scattering in ring arms~(squares in Fig.~\ref{fig6}(a)) provides the necessary total flux phase upon encircling the ring, i.e. traversing both arms. We present the ballistic case for the upper arm:
\begin{equation}
\label{scattmat}
\Su_{\frown}=e^{i\phi}\begin{pmatrix}
0 & 0 & t e^{i\pi(\Phi+d)} & a e^{i\pi\Phi} \\
0 & 0 & a e^{i\pi\Phi} & t e^{i\pi(\Phi-d)} \\
t e^{-i\pi(\Phi+d)} & a e^{-i\pi\Phi} & 0 & 0 \\
a e^{-i\pi\Phi} & t e^{-i\pi(\Phi-d)} & 0 & 0 \\
\end{pmatrix},
\end{equation}
with $t\equiv\sqrt{1-\gamma^{2}}$, $a\equiv i\gamma$ with $\gamma\in[0,1]$, and $\phi$ is an effective phase encoding for the point disorder. The probability of transmission in the same valley, $|t|^{2}$, and the probability of transmission with scattering to the opposite valley, $|a|^{2}=\gamma^{2}$ are parametrized by $\gamma$, whose value $0$ corresponds to infinite intervalley scattering time for propagating through the arms. For the lower arm we then take $\Su_{\smile}=\Su_{\frown}(\Phi\rightarrow-\Phi,d\rightarrow-d)$, as traveling from left to right must give opposite phase contributions in the two arms. The magnetoconductance curve is then calculated by the Landauer--B\"uttiker formula~\cite{Buttiker} $G(\Phi)=Tr(\tu(\Phi)\tu^{\dagger}(\Phi))$, where $\tu$ belongs to the total scattering matrix of the device, obtained by combining the ingredients we listed above. The matrix elements of the $\tu(\Phi)$ matrix that determine the magnetoconductance are obtained by explicitly solving for the outgoing amplitudes in the right terminal of the device, after fixing the incoming amplitudes in the left terminal to $1$. Let us finally explicate some symmetry constraints on the scattering matrices. The unitarity of scattering implies $\Su^{\dagger}\Su=\openone$, expressing that particle current is conserved. Time reversal plays an important role in what follows, and it implies that for any matrix $\Su$ (for a certain choice of phase relation between incoming and outgoing modes)
\begin{equation}
\label{STRS}
\Su(\Phi)=X\;\Su^{T}(-\Phi)\;X,\;\;\;\;
X=\begin{pmatrix}
0 & 1 & 0 & 0 \\
1 & 0 & 0 & 0 \\
0 & 0 & 0 & 1 \\
0 & 0 & 1 & 0 \\
\end{pmatrix},
\end{equation}
where the matrix $X$ exchanges the valleys. Valleys act in a similar way as spins, with the spin up and down modes behaving similarly under time reversal. We will come back to this issue in the concluding section.

\begin{figure}[t]
\includegraphics[width=0.85\linewidth]{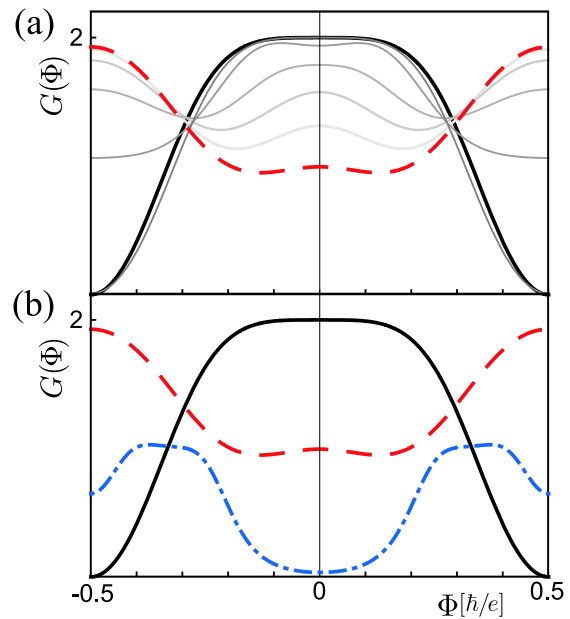}
\caption{Intervalley and disorder scattering dependent magnetoconductance at zero temperature. (a) Consider a fixed disorder configuration~($\phi=2.3$): at infinite intervalley scattering time~($a=0$ in Eq.~\eqref{scattmat}), we obtain the thick black line in the absence of a dislocation, and the thick dashed red line in presence of a $d=1/3$ dislocation. Notice the amplitude relation described in Sec.~\ref{secII'}. The thin gray lines show the evolution of the dislocated case with shortening intervalley scattering time. In the limit of maximal scattering~($a\rightarrow1$), the thick black line is reached, as if no dislocation is present. (b) Illustration of the influence of point disorder, with no intervalley scattering. The solid black line (absence of dislocation) and the dashed red line (in presence of dislocation), at a fixed disorder configuration $\phi=2.3$, are identical to the ones in part (a). The dot--dashed blue line is obtained in the presence of a dislocation, but with the disorder configuration changed to $\phi=0.1$. In contrast to the case of the red line, the blue line has no ``halving the first harmonic amplitude'' relationship~(see text) to the black line, as different disorder configurations can produce dramatically different AB oscillations.}
\label{fig3}
\end{figure}

Let us now discuss the characteristic features of the experimentally observable conductance $G(\Phi)$. The intervalley scattering $\gamma$ is the important parameter and we first analyze the case when it vanishes. This corresponds to the case of Eq.~\eqref{scattmat} after setting $\gamma=0$. It is obvious that the dislocation pseudoflux just adds to the magnetic flux. Furthermore, the two valleys are decoupled in the whole device, implying that the two currents can just be added. We can then repeat the simple argument from the beginning of this section to obtain the ``halving of amplitudes'' rule. This is independent of the particular point scatterer distribution, parametrized by the phase $\phi$ in~\eqref{scattmat}. In Figure~\ref{fig3}(a) we show the magnetoconductance without--~(thick solid line) and with~(red dashed line) a non--trivial dislocation present in the ring, with one fixed disorder phase $\phi=2.3$, where one immediately discerns the main effect of the dislocation: the fundamental harmonic is multiplied by a factor $-\frac{1}{2}$.

This example however hides a problem. Namely, a ring in the absence of dislocations, with a fixed disorder realization~($\phi\equiv\phi_{1}$), and a dislocated ring with a different disorder realization~($\phi\equiv\phi_{2}$)~(black solid and blue dot--dashed lines, respectively, in Fig.~\ref{fig3}(b)), produce different outcomes, and it becomes impossible to recognize a relationship between the two. The problem is that point disorder by itself can change the harmonic content of the AB oscillations in arbitrary ways. This has the effect that the specific information associated with the presence of the dislocation becomes completely hidden for the experimentalist, who has to produce a new sample to compare a dislocated-- with a non--dislocated AB ring, thereby changing the disorder configuration.

\begin{figure}[t]
\includegraphics[width=0.95\linewidth]{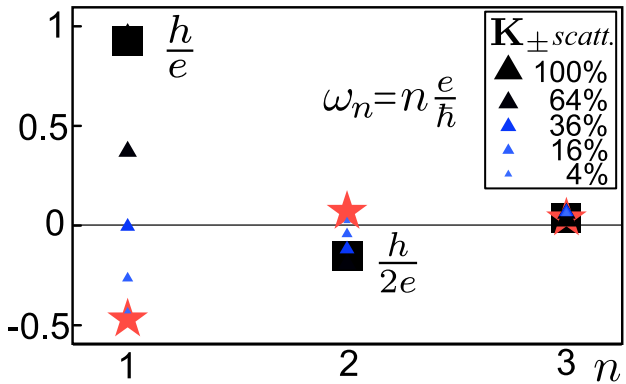}
\caption{The distributions of the disorder averaged amplitude of the harmonics of the magneto--oscillations for the ring device of Fig.~\ref{fig1} at zero temperature, with the valley polarizers switched off. The black squares indicate the response in the absence of a dislocation, and the red stars show what occurs in the presence of a $d=\frac{1}{3}$ dislocation and no intervalley scattering: the amplitudes of the fundamental and first harmonic are halved and their signs reversed~(see main text). The triangles indicate the evolution when the amount of intervalley scattering (parametrized by the value of $|\gamma|^{2}=a^{2}$, the probability of scattering between valleys on a ring arm traversal, expressed in percents) in the arms is increased.}
\label{fig4}
\end{figure}

However, the simple rule of halving the amplitude, described above, is rooted in topology, and it does survive when the point disorder is averaged over, which is a procedure that can be implemented in practice. This fact is demonstrated in Fig.~\ref{fig4}, where we show the results for the amplitudes of conductance harmonics, obtained after an averaging over the disorder phase $\phi$; the first and second harmonic amplitude of the dislocated ring (red star) have half the value (and opposite sign) compared to the ones of the ideal ring (black square).

The effect of intervalley scattering can be studied by switching on the $a$ parameter in Eq.~\eqref{scattmat}. As an illustration we show by thin gray lines in Fig.~\ref{fig3}(a) the change in magnetoconductance as we gradually decrease the intervalley scattering length; it interpolates between the outcomes of the ring with-- and without-- the dislocation. In Fig.~\ref{fig4} we show the evolution of the disorder phase averaged Fourier components, and these examples make it immediately clear that as the intervalley scattering length becomes smaller than the ring size, information regarding the presence of the dislocation is wiped out completely. The physical reason is simple. Consider again the Feynman paths; the quantum conductance is governed by paths that encircle the ring many times, and such a long path will cross the dislocation ``Dirac string'' many times. But when the intervalley scattering length is short it will randomly carry a $\bKp$ or $\bKm$ valley identity when it crosses the Dirac string, thereby picking up randomly the plus and minus dislocation Berry phase, with the obvious outcome that the net phase will average away, and this means in turn that the current will lose all information regarding the presence of the dislocation.

Finally, what is the specific effect of adding valley filters at the leads in the quantum regime? The scattering matrix describing the filter~(half black rectangle in Fig.~\ref{fig6}(a)) perfectly transmits $\sigma=+$ modes from left to right, and so time reversal symmetry fixes the form
\begin{equation*}
\Su_{pol}=\begin{pmatrix}
0 & 1 & 0 & 0 \\
0 & 0 & 0 & 1 \\
1 & 0 & 0 & 0 \\
0 & 0 & 1 & 0 \\
\end{pmatrix};
\end{equation*}
We already emphasized in the introduction that time reversal symmetry implies that the backside of a perfect valley polarizer acts like a perfect intervalley scatterer, as is further illustrated in the inset of Fig.~\ref{fig1}. Even in the absence of any other source of intervalley scattering, this implies that valley currents are no longer conserved, since the long Feynman paths will necessarily explore the backsides of the valley filters. This means that the dislocation Berry phase gets scrambled, as in the case of random intervalley scattering, and there is no simple rule for disentangling the dislocation from the non--topological random disorder. At the same time, $G(\Phi)$ will be even under all circumstances, since there is an infinity of long paths in both valleys, and B\"uttiker's theorem is obviously applicable to this case.

\section{Modeling the decoherence at finite temperature}
\label{secIII}

We can now turn to the puzzle announced in the Introduction: what happens in our device at finite temperatures? It appears that our device might represent a particular challenge to the incomplete understanding of the relation between the coherent quantum transport at short scales and classical transport at macroscopic scales that is characteristic for any system at a finite temperature. The sharpest way to express these matters is by realizing that at sufficiently large length-- and time--scales any electron system will be governed by the same hydrodynamical principles as the classical electron plasma of the high temperature limit. In contrast to the zero temperature quantum case, this classical transport is dissipative and for a Fermi--liquid the dissipation mechanisms seem well understood; they are the usual electron--electron and electron--phonon scattering lore. One can just take the Kubo formalism from the textbooks~\cite{AGD} and compute the diagrams. The problem is that such a computation becomes unmanageable for a device problem such as ours.

The argument presented in the introduction for the unevenness of the magnetoconductance at finite temperature rests implicitly on the Feynman path~\cite{mesotrans} intuition. In the first subsection we will analyze this in more detail, discovering that the argument actually rests on an uncontrolled assumption: to find out what happens with the quantum interferences at finite temperature one just sums over worldlines up to a maximal length equal to the phase coherence length, assuming that the remainder merely contributes to the incoherent current. In this way, when worldlines become ``too long'', they are assumed to just disappear. In reality these of course do not disappear but they turn into the self--energy graphs coming from the quasiparticle interactions -- the ``Kubo brick wall''. With this assumption, we obtain the finite temperature uneven magnetoconductance~(Fig.~\ref{fig5}), which becomes even at zero temperature as it should (Section~\ref{secIIIA}).

Although it is far from obvious why the cutting of world lines approach to dephasing can lead to faulty conclusions regarding the ``quantum arrow of time'', precedents exist where the intuition based on Feynman paths turned out to be misleading.~\cite{QDotInRing} It is a standard practice in mesoscopic physics to use the scattering matrix approach also at finite temperatures, and to account for the effects of dephasing using the voltage probe method invented by B\"uttiker~\cite{ButtikerPhantom}(Section~\ref{secIIIB}). Despite its simplicity and track record~(e.g., Refs.~\onlinecite{QDotInRing,ImagandPhant,exptestbutt,exptestbutt2,PhantvsImagSymm}), and the fact that by construction it respects the basic symmetries of quantum scattering, it is surely not a divine solution. A problem of principle is of course that this language, describing interfering quantum mechanical waves, is not quite the preferred way to describe finite temperature dissipative flows of classical hydrodynamics, where the quantum unitarity condition is replaced by the weaker current conservation demand. In fact, the B\"uttiker dephasing reservoirs model the effects of inelastic scattering by an effective {\em elastic} scattering.~\cite{dephaseSCBA}

For relaxational, classical hydrodynamics, time is at the heart of matter. Dealing with a problem like ours, where there are subtle complications associated with time, can the standard approach be trustworthy in the cross--over regime? We favor experimental advances in this regime. As we will show in Section~\ref{secIIIB}, the B\"uttiker construction insists that the magnetoconductance should stay even in all circumstances, even when imposing different decoherence rates for the two valleys, as is generally expected of this formalism. On the other hand, in the high temperature regime the transport turns classical, and the expectation of evenness, observed in the large body of existing experiments, is theoretically supported if given that microscopic reversibility can be viewed as certain assumptions on the classical fluctuation correlations.~\cite{Casimir,Onsager,CasOns}

\subsection{The Feynman path approach}
\label{secIIIA}

\begin{figure}[t]
\includegraphics[width=1.0\linewidth]{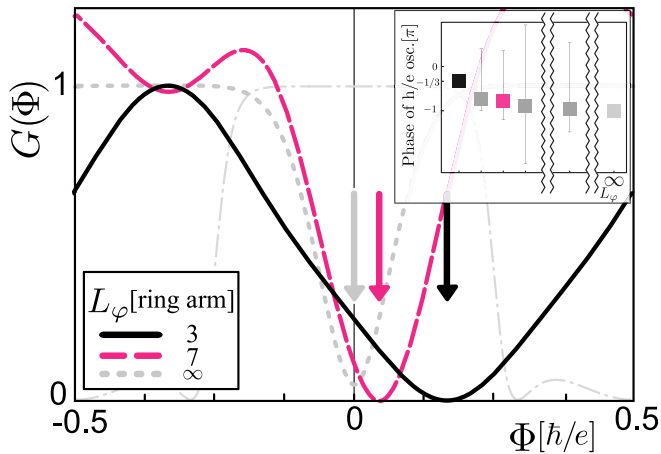}
\caption{The magnetoconductance oscillations $G(\Phi)$ as a function of applied flux $\Phi$, calculated using the ``truncated Feynman path method'' discussed in Section~\ref{secIIIA}, for the device of Fig.~\ref{fig1} with a dislocation of class $d=\frac{1}{3}$. We show the results for phase coherence lengths $L_\varphi=3,7,\infty$ in units of the ring arm length, finding that the extremum shifts from $\approx\frac{1}{3}$ the flux quantum value at high temperatures, to the origin at zero temperature. The thin dot--dashed line shows the result without a dislocation. The inset shows the range of ``disorder''~(phase $\phi$ of Section~\ref{secII'}) dependent phase shifts of the fundamental~($\frac{h}{e}$ period) harmonic of $G(\Phi)$ as a function of $L_\varphi$. Symbols show the average over the ``disorder'' phase.}
\label{fig5}
\end{figure}

We describe here the Feynman path approach explicitly.~\cite{mesotrans} We ignore intervalley scattering of any kind ($\gamma\equiv0$ after Eq.~\eqref{scattmat}), except at the polarizers, and focus on the regime where the phase coherence length $L_\varphi$ is of order of the device dimension $L$. The conductance is proportional to the electron transmission probability $\Sigma$, expressed in terms of Feynman amplitudes $\mathcal{A}_{a}$ as
\begin{equation}
\label{FP}
\Sigma=|\mathcal{A}_1+\mathcal{A}_2+\ldots+\mathcal{A}_{N(L_\varphi)}|^2+B,
\end{equation}
where we assume that in the coherent part only paths with a length not exceeding $L_\varphi$ are to be included. The longer paths contribute incoherently to the current through the term $B$, i.e. they do not produce interference terms responsible for the Aharonov--Bohm oscillations. This is the core of the dephasing model of this subsection. Let us first discuss the qualitative picture. As already explained, the perfect valley polarizer acts by being fully transparent to, say, a $\bKp$ mode propagating from left to right, and a $\bKm$ mode propagating in the opposite direction. But microscopic time reversal invariance in combination with charge conservation implies that an incoming $\bKm$ mode moving to the right is fully reflected into a $\bKp$ mode moving to the left, and vice versa~(inset Fig.~\ref{fig1}). Let us now consider the shortest possible paths that can give rise to interference in the presence of a dislocation. For a current flowing from left to right, the valley polarizers ensure that it is entirely carried by $\bKp$ modes. The current in the ``lower'' arm has to traverse the Volterra cut acquiring the phase jump while in the ``upper'' arm it is unaffected~(see Fig.~\ref{fig1}), with the net result that the transmission amplitude picks up the dislocation pseudoflux of $\frac{1}{3}$, which can in turn be compensated by an external field. Repeating the argument for a current from right to left~(dashed lines in Fig.~\ref{fig1}), one ends up with a shift of $-\frac{1}{3}$ of a flux quantum. The conclusion is that the extremum of the magneto--oscillations shifts away from its position at zero external flux, thus violating B\"uttiker's theorem. The effect is due to the finite temperature and the implicitly dissipative measurement setup, and it is present even though we consider the system very close to equilibrium.

In the present context based on formula~\eqref{FP}, the microscopic time reversal symmetry, which recovers B\"uttiker's law, is associated with the requirement that the perfect valley polarizer is a unitary intervalley scatterer for the electrons coming in with the wrong polarization. At the ``high'' temperatures discussed in the previous paragraph, the phase coherent electron encounters the polarizers at most once; it is transmitted with no opportunity to explore the ``backside'' of the polarizers. However, as the temperature is decreased one has to take into account longer and longer paths. A typical path in this ideal device is of the kind that, say, a $\bKp$ particle having traveled from left to right in the upper and lower arms, travels further via the upper arm after getting scattered to the $\bKm$ valley at the backside of the left polarizer. Such long paths destroy the valley quantization and the phase associated with the dislocation pseudoflux gets averaged away. At zero temperature paths of arbitrarily long length dominate and the extremum of the magnetooscillation obeys B\"uttiker's theorem.

To find out what exactly happens in this model as a function of decreasing temperature~(and in non--ideal devices), we computed the magnetoconductance by summing all Feynman paths with a length bounded by $L_{\varphi}$, i.e. the part of $\Sigma$ without $B$ in Eq.~\eqref{FP}. The results are shown in Fig.~\ref{fig5}, as a function of the two parameters, the dephasing length $L_\varphi$, and the disorder phase $\phi$ of Eq.~\eqref{scattmat}. Unsurprisingly, we find a smooth evolution where the extremum shifts from a flux $\approx\frac{1}{3}$ at ``high'' temperature, back to the origin as the phase coherence length increases. Only at precisely zero temperature is B\"uttiker's law recovered, since at any finite temperature the sum is always ``dominated'' by the short paths, and for this reason the effect seems quite robust.

The details of the computation go as follows: given the finite coherence length $L_\varphi$ measured in ring arm lengths, the sum over Feynman paths limited by $L_\varphi$ is performed elegantly by a simple trick. We weight the scattering matrices of the~(single transversal mode) ring arms with an auxiliary variable $\alpha$, essentially scaling $t\rightarrow\alpha t$ in Eq.~\eqref{scattmat}. Then the total coherent transmission amplitude $\mathcal{A}(\alpha,\Phi)$ is calculated exactly in the way described in Section~\ref{secII'}, by considering the scattering matrices of ring arms and polarizers that connect the various in and outgoing amplitudes in both valleys, and solving for the outgoing amplitudes.~\cite{normmetring} The advantage of doing things this way comes from the fact that every Feynman path amplitude is exactly a product of scattering matrix elements of ring arms, polarizers, and the terminals, which are accumulated as the path is followed from start to end.~\cite{mesotrans} Crucially, this implies that the amplitudes of the paths having length of $n$ ring arms (traversing an arm $n$ times) will pick up the factor $\alpha^{n}$, since a single factor $\alpha$ is associated with every pass through an arm. The sum of Feynman amplitudes $\mathcal{A}^{(n)}(\Phi)$, corresponding to paths of length up to $n$ ring arms, represents the part of the total amplitude $\mathcal{A}(\alpha,\Phi)$, where $\alpha$ appears multiplied by itself not more than $n$ times. This can be obtained by using a truncated Taylor expansion in the variable $\alpha$, because it is an expansion in terms of powers of $\alpha$, exactly what is needed. It follows that $\mathcal{A}^{(n)}(\Phi)\equiv \mathcal{A}_0(\Phi)+\alpha \mathcal{A}_1(\Phi)+\ldots+\alpha^n \mathcal{A}_n(\Phi)\Big\arrowvert_{\alpha=1}$, where we used the definition $\mathcal{A}_m(\Phi)=\frac{1}{m!}\frac{\partial^m}{\partial\alpha^m}\mathcal{A}(\alpha,\Phi)$. This transmission amplitude then gives the conductance associated with paths traversing the arms not more than $n$ times through the standard relationship $G_{coh}^{(n)}(\Phi)\sim |\mathcal{A}^{(n)}(\Phi)|^2$.

\subsection{The valley dependent B\"uttiker dephasing probe}
\label{secIIIB}

\begin{figure}[t]
\includegraphics[width=0.7\linewidth]{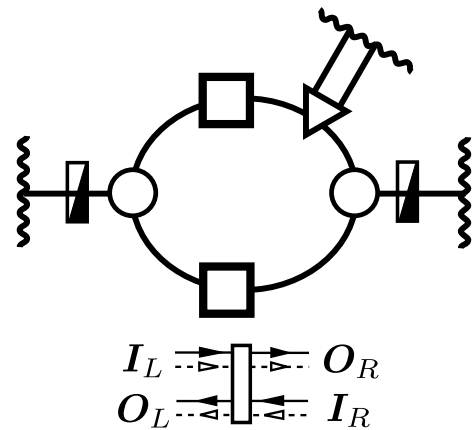}
\caption{(a) The network of scatterers modeling the Aharonov--Bohm device, with dephasing included. Wavy lines represent reservoirs and smooth lines represent wires carrying the~($\pm$) modes. The triangle element and its reservoir belong to the B\"uttiker dephasing probe construction, and are used only in Sec.~\ref{secIIIB}; the currents $I_{3}$ and $I_{3'}$ of Eq.~\eqref{curr} are flowing in the two leads connecting the triangle to its reservoir. Note that we use different chemical potentials~($\mu_{3}^{\pm}$'s) for the two valleys in this reservoir. (b) Labeling of incoming/outgoing modes for a generic scatterer, with $\Il$ representing the column $(\Il^{+},\Il^{-})$, etc.; the full/dashed lines depict the $+/-$ modes.}
\label{fig6}
\end{figure}

Let us now turn to the scattering matrix theory at finite temperature for our device by employing the ``B\"uttiker phantoms'' to model the effects of dephasing (see Fig.~\ref{fig6}).

We define $T_{pq}^{\sigma\sigma'}\equiv|\Su_{pq}^{\sigma\sigma'}|^{2}$, the modulus squared of the device scattering matrix elements, where $p,q$ refer to the leads~(terminals), and $\sigma,\sigma'\in\{+,-\}$ to the propagating modes in the leads, such that they represent the probability of scattering from mode $\sigma'$ in lead $q$ to mode $\sigma$ in $p$. It follows that the total current in lead $p$ carried by $\sigma$ electrons is
\begin{equation}
\label{Ips}
I^{\sigma}_{p}=\frac{e}{h}\sum_{q,\sigma'}T_{pq}^{\sigma\sigma'}(\mu^{\sigma}_{p}-\mu^{\sigma'}_{q}),
\end{equation}
where we use the most general option of having a different chemical potential $\mu_{p}^{\sigma}$ for each type $\sigma$ of electrons, in the reservoir connected to the $p$ lead. Such a possibility is clearly applicable when we interpret the mathematical model as describing a spin system (with two valleys being the spin up and down), while for graphene it could be less clear what different chemical potentials $\mu^{+}$, $\mu^{-}$ actually signify. In particular, one could argue that although the voltage probe is in a sense a mathematical construction that enables us to incorporate decoherence in the elastic model, it is also an actual component regularly used in the laboratory, therefore having a physical meaning. In the case of graphene, the special voltage probe would amount to having different chemical potentials at the two points in the Brillouin zone, which is conceptually conceivable. As will be elaborated below, the physical demand for equal dephasing lengths for the two valleys leads to $\mu^{+}=\mu^{-}$, and removes the problem for that situation. In any case, we regard that, conceptually, the literal interpretation of the dephasing reservoir as a physical entity is not necessary.~\cite{dephaseSCBA}

The B\"uttiker voltage probe method~\cite{ButtikerPhantom} is based on the idea that electrons lose their phase in reservoirs, thus one extends the system by introducing $N-2$ additional, auxiliary (``phantom'') reservoirs (labeled by $\tilde{f}\in\{\tilde{3},\tilde{N}\}$), where every one of them is coupled to the device through two familiar leads~(labeled $f$ and $f'$ at reservoir $\tilde{f}$), each carrying the two~(``$\pm$'') modes, but with the constraint that the total current towards a reservoir $I_{\tilde{f}}\equiv0$, i.e. the reservoirs will not drain current, but will provide dephasing. The choice of two leads (instead of, e.g. one) is just to make possible total decoherence.~\cite{ButtikerDeph} Effectively, one solves these $N-2$ current constraints (linear equations) for the {\em a priori} unknown $N-2$ auxiliary phantom chemical potentials $\mu_{\tilde{f}}$, and then eliminates these $\mu_{\tilde{f}}$ in the expressions for the currents in the physical leads. Performing this elimination in the physical current equations leads to new, effective transmission coefficients between the physical leads which figure in these equations. These effective transmission coefficients are then functions of the extended system's transmission coefficients between the physical leads, as well as the transmission coefficients to the phantom leads. To recall the familiar results of Ref.~\onlinecite{ButtikerPhantom}, let us briefly specialize to the simplest, single mode, two terminal case, dropping thereby the $\sigma$ index, as well as having the simple expression for the conductance $G=T_{12}$ implied by Eq.~\eqref{Ips}. Then for example in the case of one phantom lead ($\tilde{f}=3$), the above elimination procedure yields,
\begin{equation}
\label{phantom}
T_{12,eff}=T_{12}+\frac{T_{1\tilde{3}}T_{\tilde{3}2}}{1-T_{\tilde{3}\tilde{3}}},
\end{equation}
where $T_{\tilde{3}p}$ is to be understood as the total transmission coefficient from lead $p$ to the dephasing reservoir $\tilde{3}$, e.g. $T_{\tilde{3}1}=T_{31}+T_{3'1}$, etc. The form of the conductance $G_{eff}=T_{12,eff}$ obtained in this way tells us that the current divides into a coherent ($T_{12}$), and an incoherent piece, where the second term is quite suggestive: electrons starting from lead $2$ leave the device to $\tilde{3}$ and come back to lead $1$, while we have to multiply the probabilities to obtain the answer --- the classical, incoherent way of propagation. The amount of decoherence is determined by the probability of scattering into the dephasing leads. For instance, when electrons leave into the dephaser with unit probability, the coherent contribution to the net conductance vanishes completely, since unitarity of scattering requires that $T_{p\tilde{3}}=1$, $p\in\{1,2\}$ implies $T_{12}=0$ in Eq.~\eqref{phantom}. The bottomline is that the effects of inelastic scattering are mimicked by a model system of elastic scatterers with extra leads added, while the current constraints become nonlinear in the amplitudes (linear in their moduli squared), thereby scrambling the phase information.

An appealing feature of this method is that the effective system including the decoherence automatically respects the symmetries of the original scattering problem,~\cite{ButtikerSymm} in so far as it is encoded in the $T$ matrices. One can explicitly check by using formulas of the variety~\eqref{phantom} (see Ref.~\onlinecite{ButtikerSymm}), that unitarity (sum of elements of any row or column of $T$ equals $1$), and time reversal symmetry ($T_{pq}(\Phi)=T_{qp}(-\Phi)$), of the starting extended $T$ matrix imply precisely the same symmetries for the $T_{eff}$ matrix. These two symmetries are sufficient to derive B\"uttiker's theorem on the evenness of the magnetoconductance for a two terminal device.~\cite{Buttiker}

We are now ready to address our graphene device. The essential ingredient is the demand that valley currents be conserved, such that the Berry phase of the dislocation becomes active. The implication is that the phantom reservoirs have to respect valley conservation. This is at odds with the notion of an equilibrium reservoir that would back inject valley currents with equal probability regardless the nature of the current it swallows. The standard dephasing reservoirs of the B\"uttiker theory are obviously of this equilibrium kind and we have to modify the construction to do justice to the conservation law associated with the ``internal'' valley quantum number. We first emphasize again that in the standard treatment of the single mode case~\cite{ButtikerPhantom} with one dephasing probe~(labeled $\tilde{3}$), one imposes the hydrodynamical conservation of the total current by setting $I_{\tilde{f}=3}=0$, which then leads to Eq.~\eqref{phantom} after elimination of the chemical potential $\mu_{\tilde{3}}$. In order to allow a maximal current flow in and out of the dephasing reservoir, one equips it with two leads labeled by $f=3,3'$ as we already discussed. Thereby the hydrodynamical current conservation turns into the constraint
\begin{equation}
I_{3}+I_{3'}=0,
\end{equation}
requiring that the dephasing reservoir drains no net current. The scattering matrix connecting the two physical leads and the two phantom leads can be symbolically represented by a triangle as in Fig.~\ref{fig6}.

In order to impose the crucial valley current conservation as well, we now generalize this construction by setting
\begin{equation}
\label{curr}
\begin{aligned}
I_{3}^{+}+I_{3'}^{+}&=0, \\
I_{3}^{-}+I_{3'}^{-}&=0.
\end{aligned}
\end{equation}
In this way we enforce that the decoherence happens independently for the two valley currents.

It is obvious that we have to introduce two chemical potentials, $\mu_{3}^{+}$ and $\mu_{3}^{-}$, and use them to enforce the two constraints in Eq.~\eqref{curr}. Since $\mu_{3}^{+}$ and $\mu_{3}^{-}$ can be used as independent parameters, one may conclude that the intrinsic non--equilibration of the conserved valley currents is expressed as a non--equilibrium state of the phantom reservoir, keeping in mind that in principle this reservoir has no physical existence --- it is just a trick to encode that electrons moving through the ring at finite temperature will dissipate their energy by exciting phonons and electron--hole pairs. In summary, the constraints in Eq.~\eqref{curr} are coding for the non--standard ingredient that an ``internal'' (valley) quantum number is conserved. Finally, we emphasize that there is an additional freedom in the choice of the scattering matrix~($\Su_d$) associated with the way the dephaser is connected to the ring: the scattering indicated by the triangle element of Fig.~\ref{fig6}. This contains the transmission coefficients into the dephaser, and thereby controls the degree of decoherence caused by the dephaser. The $\Su_d$ does not mix the valleys, so it determines the intravalley dephasing time. Obviously in the physical system the decoherence in the two valleys should be the same, leading to constraints on the matrix elements discussed in detail below. Given these ingredients, the calculations are straightforward and are summarized in the Appendix.

The outcome for magnetoconductance computed from the results in the Appendix is as follows. According to expectations, we analytically prove that the magnetoconductance $G(\Phi)$ is even in the flux $\Phi$, assuming the symmetries (time reversal and unitarity) of the $T$ matrix are present. The evenness is thus independent of the values of all physical parameters, and persists even when different dephasing lengths are assigned to the two valleys by tuning the $\Su_d$ matrix. Such a situation corresponds to a non-equilibrated reservoir, with $\mu_{3}^{+}\neq\mu_{3}^{-}$. Invariably, these chemical potentials scale with the physical voltage $\mu_{1}-\mu_{2}$ (no $\pm$ dependence in physical reservoirs), consistent with the linear response regime.

\begin{figure}[t]
\includegraphics[width=0.9\linewidth]{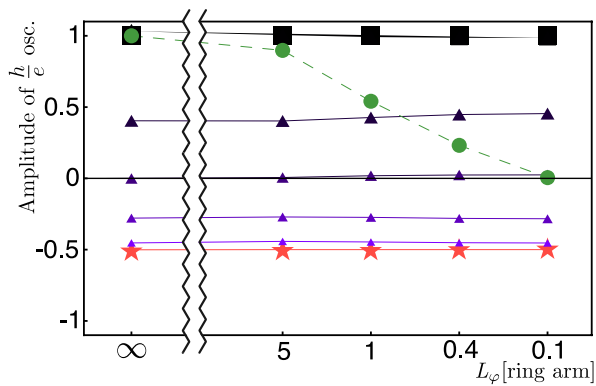}
\caption{The temperature, parametrized through the dephasing length $L_{\varphi}$, dependence of the disorder phase averaged $\frac{h}{e}$ harmonic amplitude, within the B\"uttiker dephasing model. The symbols are taken from Fig.~\ref{fig4} --- black square denoting the ideal ring, red star the dislocated ring with no intervalley scattering, and triangles the dislocated ring with varying intervalley scattering lengths. The first, $L_\varphi=\infty$ entry reproduces the zero temperature result of Fig.~\ref{fig4}. At each separate finite dephasing length value, amplitudes are normalized by the ideal ring amplitude (black square). The figure then shows how the ratio of the dislocated ring harmonic (red star) to the ideal ring harmonic varies negligibly from the zero temperature value of $-1/2$. The green dots represent the values of the ideal ring un--normalized amplitudes at each dephasing length, and show how the oscillations disappear with rising temperature.}
\label{fig7}
\end{figure}

Let us now analyze the oscillations themselves. At zero temperature, when the scattering into the phantom reservoirs vanishes, the model reduces to the matters discussed in Section~\ref{secII'}. The corresponding results for the disorder phase averaged amplitude of the fundamental, $\frac{h}{e}$ harmonic, seen as the first entry of Fig.~\ref{fig4}, are shown as the infinite dephasing length ($L_\varphi=\infty$) entry in Fig.~\ref{fig7} (black square --- in absence of dislocation, red star --- in presence of dislocation, triangles --- with dislocation and varying intervalley scattering length). The effect of finite temperature is modeled by switching on the scattering into the dephasing reservoir, and amounts to a gradual decrease of the magnetoconductance oscillations that eventually vanish when the dephasing length becomes small compared to the device dimensions: the green dashed line and green circles of Fig.~\ref{fig7} show the overall oscillation amplitude dependence on the dephasing length $L_\varphi$. The next issue is how the ratio between the disorder phase averaged Fourier amplitudes of the dislocated and ideal ring evolve with temperature. The Fig.~\ref{fig7} shows that this ratio is virtually independent of the temperature, and retains the value of $-\frac{1}{2}$ identified at zero temperature (Section~\ref{secII'}).

\section{Conclusions}
\label{secIV}

Can Casimir--Onsager relations be invalidated because leftovers of quantum coherent currents at intermediate temperatures cannot equilibrate in a true sense due to a conservation law applying to an internal symmetry? The special features of the device introduced in this paper make this provocative question germain. We do not claim to have a definitive answer. Within the realm of finite temperature quantum transport the issue appears to be unresolved and we challenge the readership to devise a more complete theoretical treatment that has the capacity to settle these matters. We hope that the considerations in this paper will motivate the experimentalists to focus in on the physics of dislocations in graphene. It seems there are no fundamental obstructions to the realization of our proposed device, with the possible exception that it might appear challenging to keep valley currents conserved on reasonable length scales. On the other hand, such an experiment still represents a considerable technical challenge, but the reward is potentiality of probing the reach of validity of a familiar law in a novel context.

We do invoke specialties of graphene, but the theme is more general. Are there other conserved internal quantum numbers that can be utilized for similar purposes? The transport of spin comes immediately to mind, with spin polarization taking the role of valley polarization, spin currents~\cite{PhysRevB.66.060404,SeebeckSpin} as valley currents, and spin--orbit coupling being like intervalley scattering. One needs more equipment. It appears that in principle the Aharonov-Casher Berry phase~\cite{seminalAC} associated with an electrical monopole in the middle of the ring has the potential to take the role of the dislocation (see, e.g., Ref.~\onlinecite{Bart}), but a literal analogue of valley polarizers is less obvious.

This brings us back to an important byproduct of this pursuit: the graphene dislocation with its Berry phase that communicates with valley currents in a unique way. More speculatively, if ``valleytronics'' ever gets off the ground, and the Casimir--Onsager relations are shown to fail in the intermediate regime (however unlikely the prospect), the dislocations would have their use as unique valleytronic circuit elements measuring in a topologically robust ways the valley polarized currents. The equipment based on valley filters, that were at the focus of this paper, might not be the best way to go, and the same objection holds for other possible microscopic mechanisms of producing valley polarized states.~\cite{DomainWall} An analogy with the quantum spin Hall effect~\cite{kane:226801,kane:146802,bernevig:106802,MarkusKonig11022007} suggests another alley to explore. Topological band insulation rooted in spin--orbit coupling goes hand in hand with chiral spin currents at the surface, and it is imaginable~\cite{PhuanOngprivcomm} that these can be exploited to construct a spin battery. It was recently argued that similar topologically protected currents exist at the interface between graphene bilayers, where the gap associated with AB sublattice breaking changes sign.~\cite{BLGtop} These chiral interface states are associated with valley currents and one can contemplate to exploit these for the purpose of constructing a valley battery.

\acknowledgments{The authors would like to acknowledge useful discussions with A. Morpurgo, J. E. Moore, F. D. M. Haldane. This work is financially supported by the Nederlandse Organisatie voor Wetenschappelijk Onderzoek~(NWO), and the Stichting voor Fundamenteel Onderzoek der Materie~(FOM).}

\section{Appendix}

We discuss here the details of the scattering matrix calculations which include the valley preserving dephasing reservoirs, as discussed in Section~\ref{secIIIB}. We follow in detail the method and interpretation introduced in Refs.~[\onlinecite{Buttiker,ButtikerSymm}]. We eliminate the dephaser, keeping everything else in the system arbitrary. At this point we need only the information that there are two time reversed modes in the system, while the dephaser takes the form~\eqref{curr}, and the full $T$ matrix respects the basic symmetry requirements, i.e.\! unitarity and time reversal. We label the eliminated dephaser by $\tilde{f}=\tilde{3}$, keeping the label $p$ for all the other leads of the device, some of which might refer to other dephasers. By applying the expression for the current in an arbitrary lead~\eqref{Ips} to the dephaser~(i.e. putting $p=3,3'$), we determine the phantom potentials $\mu_{3}^{+}$, $\mu_{3}^{-}$, which are present in the constraint equations~\eqref{curr}, in terms of the other potentials. Once these phantom potentials are determined, we eliminate them in the expressions for the currents in all the other terminals of the device~(i.e. $p\neq3,3'$ in Eq.~\eqref{Ips}). These expressions then describe the currents in all the remaining terminals, as a function of their associated chemical potentials in the form of an effective matrix $T_{eff}$ with elements $T_{eff,pq}^{\sigma\sigma'}$. The general expressions for the elements of $T_{eff}$, and the chemical potentials of the eliminated dephasing reservoir are
\begin{equation}
\label{Teff}
\begin{aligned}
T_{eff,pq}^{\s\s'}& =T_{pq}^{\s\s'}+T_{p\tilde{3}}^{\s+}M^{+\s'}_{q}+T_{p\tilde{3}}^{\s-}M^{-\s'}_{q},\\
\mu_{3}^{\pm}& =\sum_{q\not\in\tilde{3}}\sum_{\s}M_{q}^{\pm\s}\mu_{q}^{\s},\\
M_{q}^{\rho\s}& \equiv\frac{1}{\Delta}\left[(2-T_{\tilde{3}\tilde{3}}^{\bar{\rho}\bar{\rho}})T_{\tilde{3}q}^{\rho\s}+T_{\tilde{3}\tilde{3}}^{\rho\bar{\rho}}T_{\tilde{3}q}^{\bar{\rho}\s}\right],\\
\Delta& \equiv(2-T_{\tilde{3}\tilde{3}}^{++})(2-T_{\tilde{3}\tilde{3}}^{--})-T_{\tilde{3}\tilde{3}}^{+-}T_{\tilde{3}\tilde{3}}^{-+},
\end{aligned}
\end{equation}
where $\bar{\s}$ denotes the valley opposite to $\s$ ($\bar{\s}=\mp$ for $\s=\pm$), and we used the obvious abbreviations for summing over the leads ($3$, $3'$$\in\tilde{3}$), e.g. $T_{p\tilde{3}}^{\s\s'}\equiv T_{p3}^{\s\s'}+T_{p3'}^{\s\s'}$, $T_{\tilde{3}\tilde{3}}^{\s\s'}\equiv T_{33}^{\s\s'}+T_{33'}^{\s\s'}+T_{3'3}^{\s\s'}+T_{3'3'}^{\s\s'}$, etc.

One can in principle now proceed to eliminate the next dephasing circuit element, using the same procedure outlined above, of course now applied to the effective matrix $T_{eff}$ defined in Eq.~\eqref{Teff}, which describes the system at this stage. If this is repeated iteratively for all phantoms, the final effective $T_{phys}$, which describes scattering between the physical leads, is obtained. The general implication of this dephasing model, proven in the following, is that the two basic symmetries (unitarity and time reversal) hold for $T_{eff}$ in Eq.~\eqref{Teff}, which in turn implies that they will hold also for each $T$ matrix obtained by successive eliminations of dephasing reservoirs, including the sought $T_{phys}$ at the final stage. If $T_{phys}$ describes a two terminal device, we conclude that the assumptions of B\"uttiker's theorem hold, and the magnetoconductance is even. The unitarity of scattering described by $T_{eff}$, expressed as $\sum_{q,\sigma'}T_{eff,pq}^{\sigma\sigma'}=1$, can be translated into the equality $\sum_{q\notin \tilde{3},\s}M^{\rho\s}_{q}=1$, and this equality is proved correct directly by using the unitarity of the starting $T$ matrix, i.e. $\sum_{q,\sigma'}T_{pq}^{\sigma\sigma'}=1$. Time reversal symmetry of the original matrix is expressed through $T_{pq}^{\s\s'}(\Phi)=T_{qp}^{\bar{\s}'\bar{\s}}(-\Phi)$, as is implied by the time reversal property of the matrix $\Su$, Eq.~\eqref{STRS}. It is straightforward to show that this property also holds for $T_{eff}$ of Eq.~\eqref{Teff}, by checking it for explicit values of $\s$ and $\s'$, with use of the property for $T$; a simplification comes from noting that $\Delta(\Phi)=\Delta(-\Phi)$. This finishes our analysis of the general case. For the particular case of our two terminal device model with one dephasing element in the ring arm, Fig.~\ref{fig6}, we also evaluated the conductance $G_{phys}=\sum_{\s\s'}T_{phys,12}^{\s\s'}$ numerically, and the results are discussed in Section~\ref{secIIIB}. We crudely estimate the dephasing length $L_{\varphi}$, used qualitatively in Fig.~\ref{fig7}, by the simple formula $\exp{(-1/L_{\varphi})}\equiv 1-|\varepsilon|^{2}$, where $\varepsilon$ is the element of the $\Su_{d}$ matrix which describes the scattering from the ring arm into the dephasing reservoir. The value of $1-|\varepsilon|^{2}$ is then the probability for the electron not to dephase while traversing the arm. By considering each ring arm traversal (path of length $1$) as an independent statistical measurement of this probability, the characteristic dephasing length $L_{\varphi}$ follows. This is analogous to writing the formula for the half life of an unstable particle having the probability $|\varepsilon|^{2}$ to decay.

To clarify the structure of the quantities calculated above, Eq.~\eqref{Teff}, and compare them to the one mode case of Ref.~\onlinecite{Buttiker}, we now specialize to the relevant two terminal (labeled interchangeably by $L$, $R$ for ``left'' and ``right'', and $1$, $2$, respectively) device with one additional dephasing reservoir $\tilde{3}$. It will then be useful to introduce quantities which group $T$ matrix elements according to their meaning, so we define: the total probability of scattering the electron from the physical lead $p$ into a $\s$ electron exiting into the dephasing reservoir by $S_{i}^{p\s}\equiv T_{3p}^{-+}+T_{3p}^{--}+T_{3'p}^{-+}+T_{3'p}^{--}$, the probability of scattering an incoming $\s$ electron from the dephasing reservoir into the $p$ physical lead by $S_{o}^{p\s}\equiv T_{p3}^{+\s}+T_{p3}^{-\s}+T_{p3'}^{+\s}+T_{p3'}^{-\s}$, as well as the obvious quantities $S_{i/o}^{p}=S_{i/o}^{p+}+S_{i/o}^{p-}$, $S_{i/o}^{\s}=S_{i/o}^{1\s}+S_{i/o}^{2\s}$, and $S_{i/o}=S_{i/o}^{1}+S_{i/o}^{2}$. A physical quantity of relevance is the amount of energy dissipated by the dephaser, which acts as an inelastic scatterer,~\cite{Buttiker} by the $\s$ current. This describes the amount of dephasing of the $\s$ current, and we express it through a dimensionless function $\eta^{\s}$ as $W^{\pm}=\frac{1}{h}\eta^{\pm}(\mu_{L}-\mu_{R})^{2}$. It reflects the fact that the electrons exchanged between the physical and dephasing reservoir are injected at different chemical potentials. If one also defines dimensionless functions $\chi^{\s}$ to describe the phantom chemical potentials by $\mu_{3}^{\pm}=\mu_{R}+\chi^{\pm}(\mu_{L}-\mu_{R})$, and introduce one more auxiliary combination of matrix elements by $C_{p\s}\equiv S_{o}^{\bar{\s}}S_{i}^{p\s}+T_{\tilde{3}\tilde{3}}^{\s\bar{\s}}S_{i}^{p}$, then for the two terminal device it can be written that
\begin{equation}
\label{W}
\begin{aligned}
\chi^{\s}& =\frac{C_{1\s}}{\Delta},\\
\Delta& =C_{1+}+C_{2+}=C_{1-}+C_{2-},\\
\eta^{\s}& =S_{i}^{1\bar{\s}}+T_{\tilde{3}\tilde{3}}^{\s\bar{\s}}(\chi^{\bar{\s}})^{2}-(2-T_{\tilde{3}\tilde{3}}^{\s\s})(\chi^{\s})^{2},\\
T_{eff,21}& =T_{21}+S_{o}^{2-}\chi^{-}+S_{o}^{2+}\chi^{+}=T_{eff,12}.
\end{aligned}
\end{equation}
Physically, one expects the degree of decoherence to be the same for the two valleys, but we emphasize again that this is not needed for the evenness of magnetoconductance, since no such assumption was made in its proof at the beginning of the Appendix. This demand reads $W^{+}=W^{-}$, and in general puts constraints on the matrix elements. Although it is not trivial to extract the simplest condition equivalent to this demand, we note that in the case when $S_{i}^{1+}=S_{i}^{1-}=\frac{1}{2}S_{i}^{1}\equiv\underline{S}_{i}^{1}$, and $S_{i}^{2+}=S_{i}^{2-}=\frac{1}{2}S_{i}^{2}\equiv\underline{S}_{i}^{2}$ (the use of underline $S$ symbols here should of course not confuse with the same symbol for scattering matrices), which corresponds to saying that the probability of the electron coming from the first lead to scatter into the dephasing reservoir as $+$ or $-$ electron is the same, and this statement holds also for the second lead, significant simplifications occur.Namely, if we put $S_{i/o}^{+}=S_{i/o}^{-}=\frac{1}{2}S_{i/o}\equiv\underline{S}_{i/o}$, the quantities in Eq.~\eqref{W} become,
\begin{equation}
\label{Wpm}
\begin{aligned}
\chi^{+}& =\chi^{-}=\frac{\underline{S}_{i}^{1}}{\underline{S}_{i}},\\
\eta^{+}& =\eta^{-}=\frac{\underline{S}_{i}^{1}\underline{S}_{i}^{2}}{\underline{S}_{i}}=\frac{\underline{S}_{i}^{1}\underline{S}_{i}^{2}}{\underline{S}_{i}^{1}+\underline{S}_{i}^{2}},\\
T_{eff,21}& =T_{21}+2\frac{\underline{S}_{o}^{2}\underline{S}_{i}^{1}}{\underline{S}_{i}}.
\end{aligned}
\end{equation}
The two valleys contribute equally to the incoherent part of the transmission in the last equality. Finally, we note that if additionally to the above assumption, which leads to $W^{+}=W^{-}$, we assume a vanishing magnetic field, the time reversal condition $S_{i}^{2}(\Phi)=S_{o}^{2}(-\Phi)$ becomes $S_{i}^{2}=S_{o}^{2}$, and we see from~\eqref{Wpm} that the dimensionless dissipated energy becomes equal to the incoherent transmission, i.e. $2\eta=T_{eff,21}-T_{21}$. The factor $2$ accounts for two modes; this recovers the result of Ref.~\onlinecite{ButtikerPhantom}, which considered a single mode and zero magnetic field situation.

\bibliographystyle{apsrev}
\bibliography{graph}

\end{document}